\documentclass[pra, 10pt, twocolumn, letter, floatfix]{revtex4}

\usepackage{amsmath}
\usepackage{amsfonts}
\usepackage{amssymb}
\usepackage{graphicx}
\usepackage{rotating}
\allowdisplaybreaks
\usepackage{setspace}
\usepackage{bm}

\begin{document}

\title{Variation After Response in Quantum Monte Carlo}

\author{Eric Neuscamman$^{1,2,}$\footnote[1]{Electronic mail: eneuscamman@berkeley.edu}}
\affiliation{$^1$Department of Chemistry, University of California, Berkeley, California 94720, USA\\
             $^2$Chemical Sciences Division, Lawrence Berkeley National Laboratory, Berkeley, California 94720, USA}

\begin{abstract}
We present a new method for modeling electronically excited states that overcomes a key failing of linear response
theory by allowing the underlying ground state ansatz to relax in the presence of an excitation.
The method is variational, has a cost similar to ground state variational Monte Carlo, and admits both open
and periodic boundary conditions.
We present preliminary numerical results showing that,
when paired with the Jastrow antisymmetric geminal power ansatz, the variation-after-response formalism delivers
accuracies for valence and charge transfer single excitations on par with equation of motion coupled cluster,
while surpassing even this very high-level method's accuracy for excitations with significant doubly excited
character.
\end{abstract}

\maketitle



Linear response (LR) theory is currently the most widely used approach for modeling electronic excitations in molecules.
It has been applied to a wide variety of approximate wave functions
\cite{HeadGordon:2005:cis_review,Krylov:2008:eom_cc_review,Chan:2014:lrdmrg,Chan:2013:thouless,Verstraete:2013:mps_tangent},
most notably the Slater determinant (SD) within the context of time dependent density functional theory (TDDFT)
\cite{Gross:2004:tddft_review}.
Its allure is straightforward:  given a sufficiently flexible ansatz, the LR of the ground state to a time
dependent perturbation will show resonances at each of the molecule's excitation energies.
In practice, of course, we employ approximate ansatzes that are nonlinear in their variables, and so
a qualitatively correct treatment can only be expected for those excitations that lie within the ansatz's
LR space.
Even then, the very nature of LR prevents the nonlinear relaxation of the ansatz in the presence of the
excitation, which can be important for fully relaxing charge densities and correlation effects and often
leads to eV-sized errors in the predicted excitation energy
\cite{HeadGordon:2005:cis_review,Zhao:2016:eom_jagp}.
Such relaxations are especially important in charge transfer excitations
\cite{Subotnik:2011:cis_ct_bias}
and double excitations
\cite{Bartlett:1996:eom_cc_butadiene,Grimme:1999:dft_mrcis,Burke:2004:dressed_tddft},
both of which play important roles in photochemistry.
In this report, we show how these nonlinear relaxation effects can be captured through a variational 
optimization of an ansatz together with its linear response.

In the equation of motion (EOM) approach to LR theory, the Schr\"{o}dinger equation is diagonalized in a
subspace of Hilbert space spanned by the first derivatives of the ground state with
respect to its variables.
For a SD, this space contains all single excitations
\cite{HeadGordon:2005:cis_review}
(hence the name configuration interaction singles or CIS), while
for coupled cluster (EOM-CCSD) it will also contain all double excitations
\cite{Krylov:2008:eom_cc_review}.
In general, we may think of an EOM ansatz,
\begin{align}
\left|\Psi_{\mathrm{EOM}}(\vec{x},\vec{\mu})\right> = \sum_i \mu_i \left|\frac{\partial\Phi(\vec{x})}{\partial x_i}\right>
\label{eqn:eom_wfn}
\end{align}
in which $|\Phi(\vec{x})\rangle$ is the ground state ansatz with variables $\vec{x}$.
In the traditional EOM approach, $\vec{x}$ is held fixed at the values determined during the ground state optimization.
However, if our intention is to produce the best description possible for an excited state using $|\Psi_{\mathrm{EOM}}\rangle$
as our ansatz, then these values for $\vec{x}$ are not optimal.
In CIS, for example, the shapes of the closed shell orbitals do not relax in the presence of the excitation as they should,
a shortcoming that in principle could be repaired by relaxing $\vec{x}$ to satisfy some chosen criteria of optimality.

The central idea in this study will be to optimize $\vec{x}$ and $\vec{\mu}$ together using our recently introduced
variational method for excited states \cite{Zhao:2016:direct_tar}.
This method uses variational Monte Carlo (VMC) to efficiently minimize a target function $\Omega$,
which in terms of our EOM ansatz variables $\vec{x}$ and $\vec{\mu}$ can be written as
\begin{align}
\label{eqn:es_var_func}
\Omega(\omega,\vec{x},\vec{\mu})
&= \frac{\omega-E}{(\omega-E)^2 + \sigma^2} \\
E &= \frac{\langle\Psi(\vec{x},\vec{\mu})|H|\Psi(\vec{x},\vec{\mu})\rangle}
          {\langle\Psi(\vec{x},\vec{\mu})|  \Psi(\vec{x},\vec{\mu})\rangle} \\
\sigma^2 &= \frac{\langle\Psi(\vec{x},\vec{\mu})|(H-E)^2|\Psi(\vec{x},\vec{\mu})\rangle}
             {\langle\Psi(\vec{x},\vec{\mu})|  \Psi(\vec{x},\vec{\mu})\rangle}
\end{align}
For a chosen fixed $\omega$, the global minimum of this function is the Hamiltonian eigenstate whose energy
lies immediately above $\omega$ \cite{Zhao:2016:direct_tar}, and so by placing $\omega$ appropriately we may 
optimize an excited state by varying $\vec{x}$ and $\vec{\mu}$ to minimize $\Omega$.

To achieve an efficient minimization of $\Omega$, by which we mean with a cost comparable to ground state VMC
using the same ansatz, our VMC method requires that the wave function ansatz support the efficient
evaluation of the local values of the wave function's derivatives,
\begin{align}
\label{eqn:local_deriv}
d_{j,\bm{n}} \equiv \frac{\partial \langle\bm{n}|\Psi\rangle}{\partial x_j}
\end{align}
where for the present study $\langle\bm{n}|$ will be an occupation number vector in Fock space
(note that the theory we develop here is equally applicable to VMC in real space, in which case
we would replace $\langle\bm{n}|$ with the real space position eigenstate $\langle\vec{r}|$).
This requirement presents us with a challenge, as the derivative in Eq.\ (\ref{eqn:local_deriv})
will, when applied to $|\Psi_{\mathrm{EOM}}\rangle$, produce second derivatives of the
underlying ground state ansatz $|\Phi\rangle$.
Such second derivatives are not necessary for ground state VMC, and in general are much more
expensive to evaluate than the first derivatives that are used in both ground state VMC
\cite{Nightingale:2001:linear_method,UmrTouFilSorHen-PRL-07,TouUmr-JCP-08}
and EOM-VMC \cite{Zhao:2016:eom_jagp}.

\begin{figure}[t]
\includegraphics[width=8.0cm,angle=270]{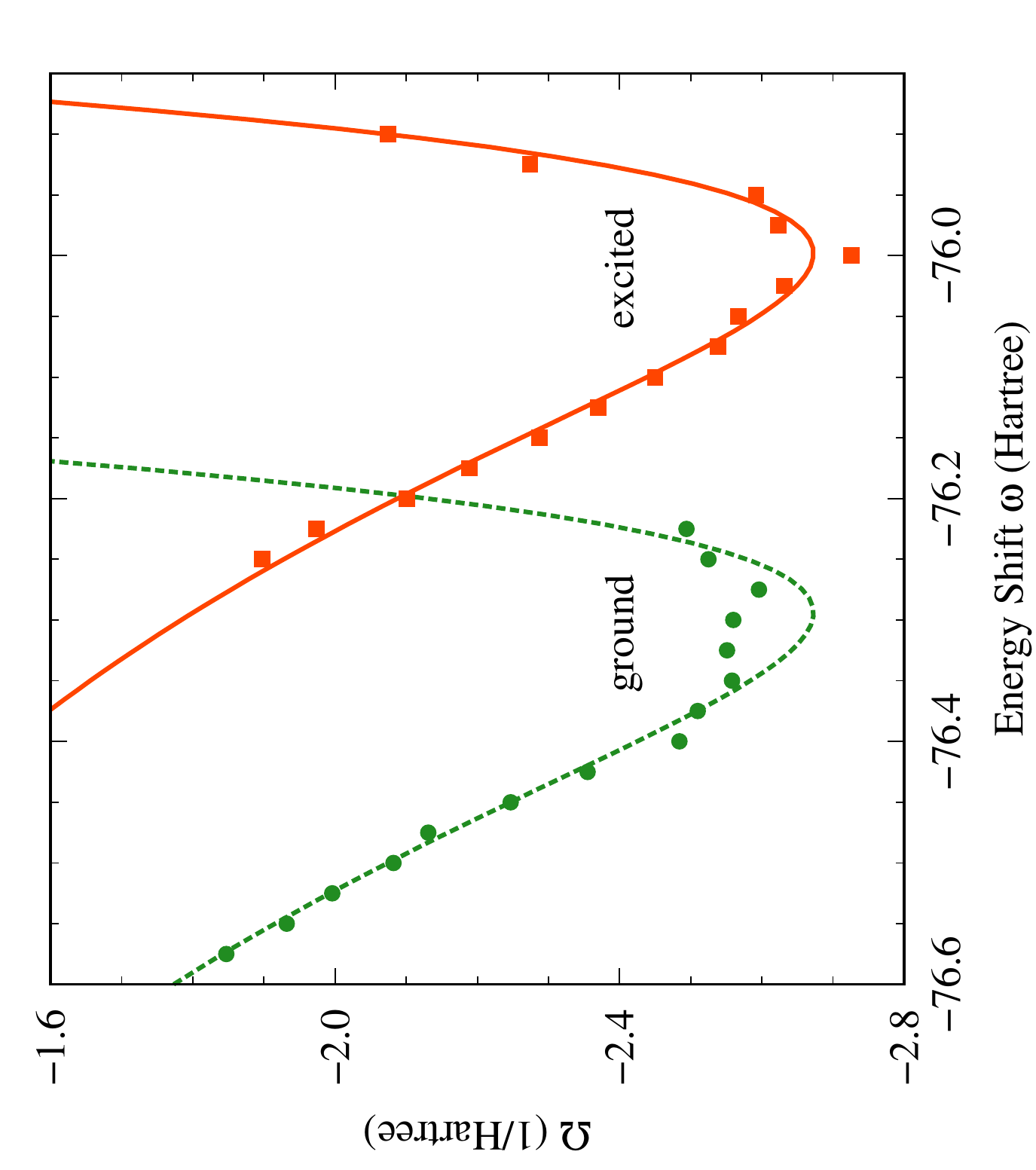}
\caption{Target function values for VAR-JAGP for the ground and first excited state of H$_2$O in a 6-31G basis.
         Points represent optimized values at individual $\omega$ while lines are fits to Eq.\ (\ref{eqn:es_var_func})
         with $\sigma^2$ and $E$ set to the variance and energy of the optimized wave function at the value
         of $\omega$ that minimizes $\Omega$.
        }
\label{fig:h2o_tf_vs_shift}
\end{figure}

To avoid the explicit evaluation of wave function second derivatives, we will therefore adopt
a finite difference approximation to our EOM ansatz,
\begin{align}
\label{eqn:var_wfn}
\left|\Psi_{\mathrm{VAR}}(\vec{x},\vec{\mu})\right> & \equiv \frac{1}{2} \Big( |\Phi(\vec{x}+\vec{\mu})\rangle - |\Phi(\vec{x}-\vec{\mu})\rangle \Big) \\
&= \left|\Psi_{\mathrm{EOM}}(\vec{x},\vec{\mu})\right> + \mathcal{O}\left(|\vec{\mu}|^2\right)
\end{align}
for which the requisite derivatives $\partial \langle\bm{n}|\Psi_{\mathrm{VAR}}\rangle/\partial x_j$ require only evaluating the
first derivatives of the underlying wave functions $|\Phi(\vec{x}+\vec{\mu})\rangle$ and $|\Phi(\vec{x}-\vec{\mu})\rangle$, as desired.
To interpret the optimization of $|\Psi_{\mathrm{VAR}}\rangle$ as variation after response (VAR), the magnitude of $\vec{\mu}$ must be small,
as otherwise the $\mathcal{O}(|\vec{\mu}|^2)$ terms will be important.
A small $\vec{\mu}$ creates no formal difficulties, as it merely leads to a small wave function norm, and in our numerical demonstrations we
have not observed any deleterious effects from the floating point arithmetic errors that one must be alert for in any finite difference approach.
Note that our method for minimizing $\Omega$ also requires derivatives of the form $\partial \langle\bm{n}|H|\Psi_{\mathrm{VAR}}\rangle/\partial x_j$
\cite{Zhao:2016:direct_tar}, but the reader may convince herself that these are, for the same reasons as for $d_{j,\bm{n}}$, no more difficult than
in the VMC's ground state linear method \cite{Nightingale:2001:linear_method,UmrTouFilSorHen-PRL-07,TouUmr-JCP-08} optimization of $|\Phi\rangle$.
In summary, by combining our VMC minimization of $\Omega$ with a finite difference representation of the EOM ansatz, we may achieve
a coupled, variational optimization of an ansatz and its linear response for an excited state for a cost similar to a VMC optimization of
the ground state.

Like our excited state variational method, the present approach is very general.
The Monte Carlo sampling may be performed in either real space or Fock space, the boundary conditions may be either open or
periodic, and the standard array of VMC ansatzes may be used.
Examples of compatible ansatzes include the Slater-Jastrow \cite{FouMitNeeRaj-RMP-01}, multi-Slater-Jastrow \cite{Morales:2012:msj},
and the Jastrow antisymmetric geminal power (JAGP)
\cite{Sorella:2003:agp_sr,Sorella:2004:agp_sr,Sorella:2007:jagp_vdw,Sorella:2009:jagp_molec,Neuscamman:2012:sc_jagp,Neuscamman:2013:jagp},
and while we look forward to investigating this wide range of opportunities in the future, we will satisfy ourselves for now
by demonstrating VAR's efficacy using the JAGP in Hilbert space.

\begin{figure}[t]
\includegraphics[width=8.0cm,angle=270]{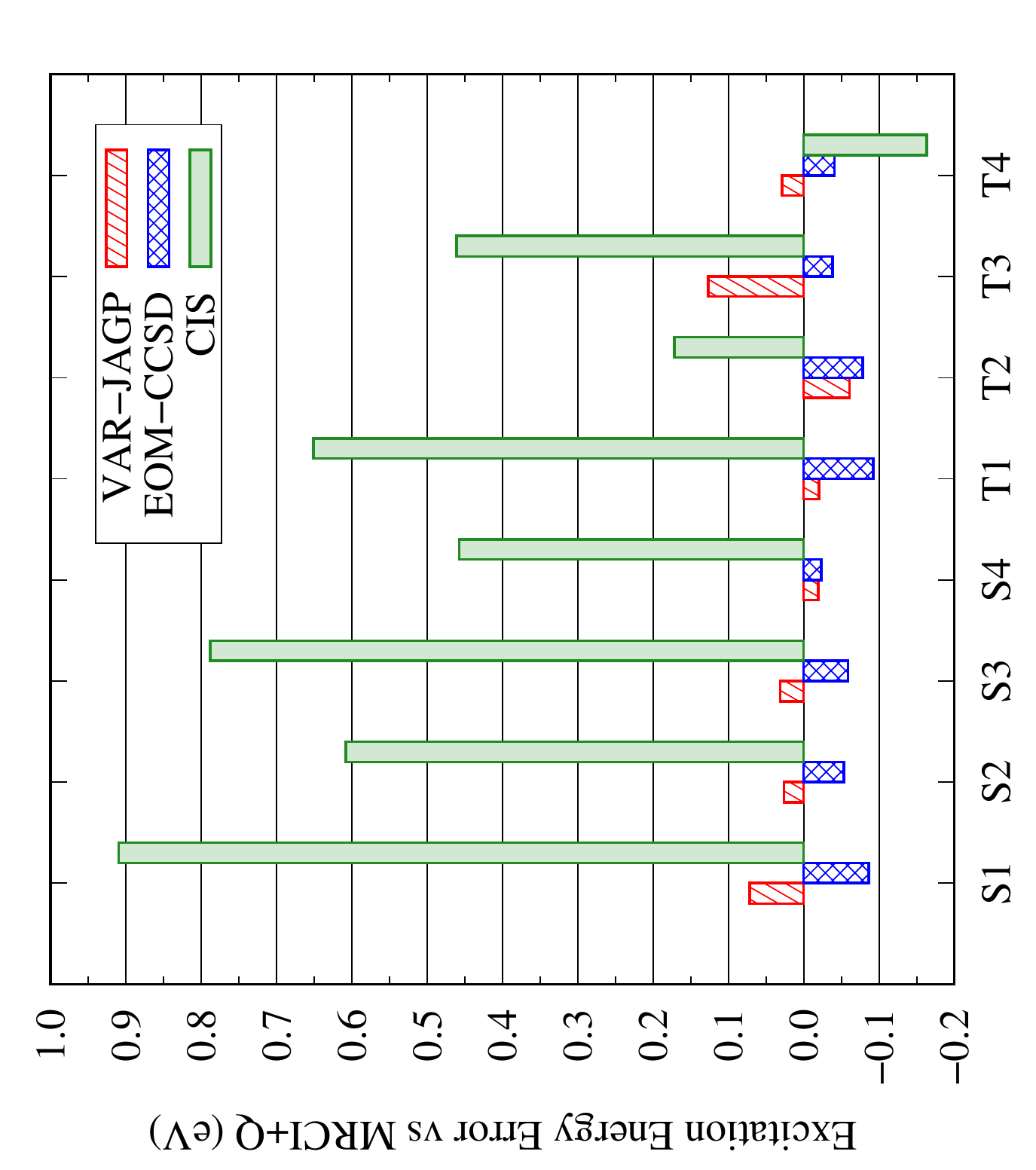}
\caption{Excitation energy errors for the first four singlet and triplet excitations in
         H$_2$O in a 6-31G basis.
        }
\label{fig:h2o_error}
\end{figure}

The JAGP ansatz in Hilbert space has been investigated previously for ground states
\cite{Neuscamman:2012:sc_jagp,Neuscamman:2013:jagp,Neuscamman:2013:cjagp}
and for excited states via both the minimization of $\Omega$ \cite{Zhao:2016:direct_tar}
and the use of a VMC-based EOM formalism \cite{Zhao:2016:eom_jagp}.
In our previous work minimizing $\Omega$ for the JAGP, we found that it can be quite challenging to choose a good initial guess for the excited state,
an issue that is easily resolved in VAR by taking the initial guess as the ground state $\vec{x}$ values and the traditional EOM solution for $\vec{\mu}$.
Indeed, some of the motivation for the present approach stems from the irony of applying EOM to a highly accurate, compact, and nonlinear
ansatz like the JAGP:  its accuracy allows for a good ground state energy, but its compactness in terms of the number of variables
leads to a small LR space and thus a large ground state bias when used in EOM \cite{Zhao:2016:eom_jagp}.
EOM-JAGP is thus, in a sense, too clever by half.
In VAR-JAGP, we expect the relaxation of $\vec{x}$ to eliminate the ground state bias and allow for excited state accuracies on par with those
achievable for ground states.

Before delving into our results, we should first address the practical issue of choosing the energy shift $\omega$.
For an exact ansatz, the energy of the state minimizing $\Omega$ will be unaffected by changes to $\omega$ except when
$\omega$ crosses an energy eigenvalue and the global minimum changes character.
For a good but approximate ansatz, we thus expect energies to be only very weakly dependent on the precise choice of $\omega$,
but it may still be desirable to make the choice unique.
This can in fact be done by minimizing $\Omega$ with respect to all its parameters, $\omega$ included, starting from an initial
guess that places one in the basin of convergence for the desired excited state.
An example of this approach is shown in Figure \ref{fig:h2o_tf_vs_shift}, where we see two distinct minima corresponding to the
ground and first excited state of H$_2$O (see Appendix for computational details).
In this case, we found that for both states, the energy $E$ changed by less than 0.02 eV when sweeping $\omega$ across a 4 eV window
centered on the $\Omega$-minimizing $\omega$, confirming that the energy is indeed quite insensitive to the precise choice of $\omega$.
In our results below, we have taken similar care to adjust $\omega$ to ensure energy insensitivity is satisfied.

\begin{figure}[t]
\includegraphics[width=8.0cm,angle=270]{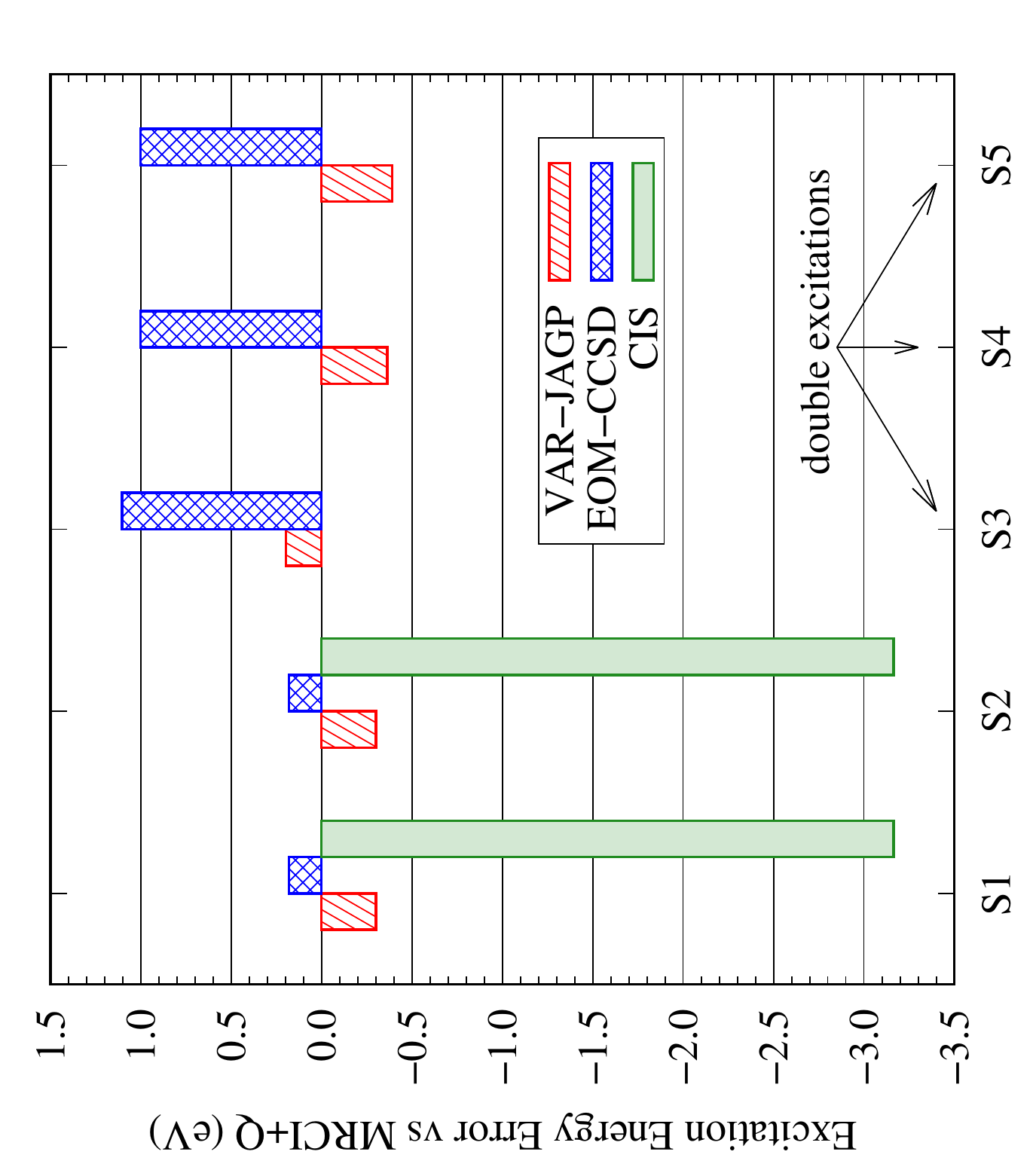}
\caption{Excitation energy errors for the first five singlet excitations in
         C$_2$ in a 6-31G basis.
        }
\label{fig:c2_error}
\end{figure}

We first tested the methodology in H$_2$O and C$_2$ by adopting the straightforward approach of using the EOM-JAGP wave functions \cite{Zhao:2016:eom_jagp}
as the initial guesses for the VAR excited state optimizations.
In Figures \ref{fig:h2o_error} and \ref{fig:c2_error}, we see that VAR-JAGP is competitive in accuracy with EOM-CCSD for single excitations,
while surpassing it in C$_2$'s low-lying double excitations.
To understand how this accuracy comes about, it is important to recognize that in a molecule with a single-reference ground state, the LR space of JAGP contains 
all single excitations and some double excitations \cite{Zhao:2016:eom_jagp}, and so we would expect it to be at least qualitatively accurate even at the EOM level
of theory.
However, as we have seen previously \cite{Zhao:2016:eom_jagp}, excitation energies in EOM-JAGP (in which $\vec{x}$ is not varied) are often much less accurate
than EOM-CCSD due to its strong ground state bias.
Here we see that VAR-JAGP eliminates this bias by relaxing $\vec{x}$, which provides for the relaxation of the orbital shapes and thus
the overall charge density and also for relaxation of correlation effects through the optimization of the Jastrow variables in $\vec{x}$.
For single excitations, these relaxation effects are also present in EOM-CCSD through its doubles operator.
For double excitations, however, EOM-CCSD's accuracy suffers as it is unable to provide these relaxations, while the VAR formalism
ensures they will still occur in VAR-JAGP.

\begin{figure}[t]
\includegraphics[width=8.0cm,angle=270]{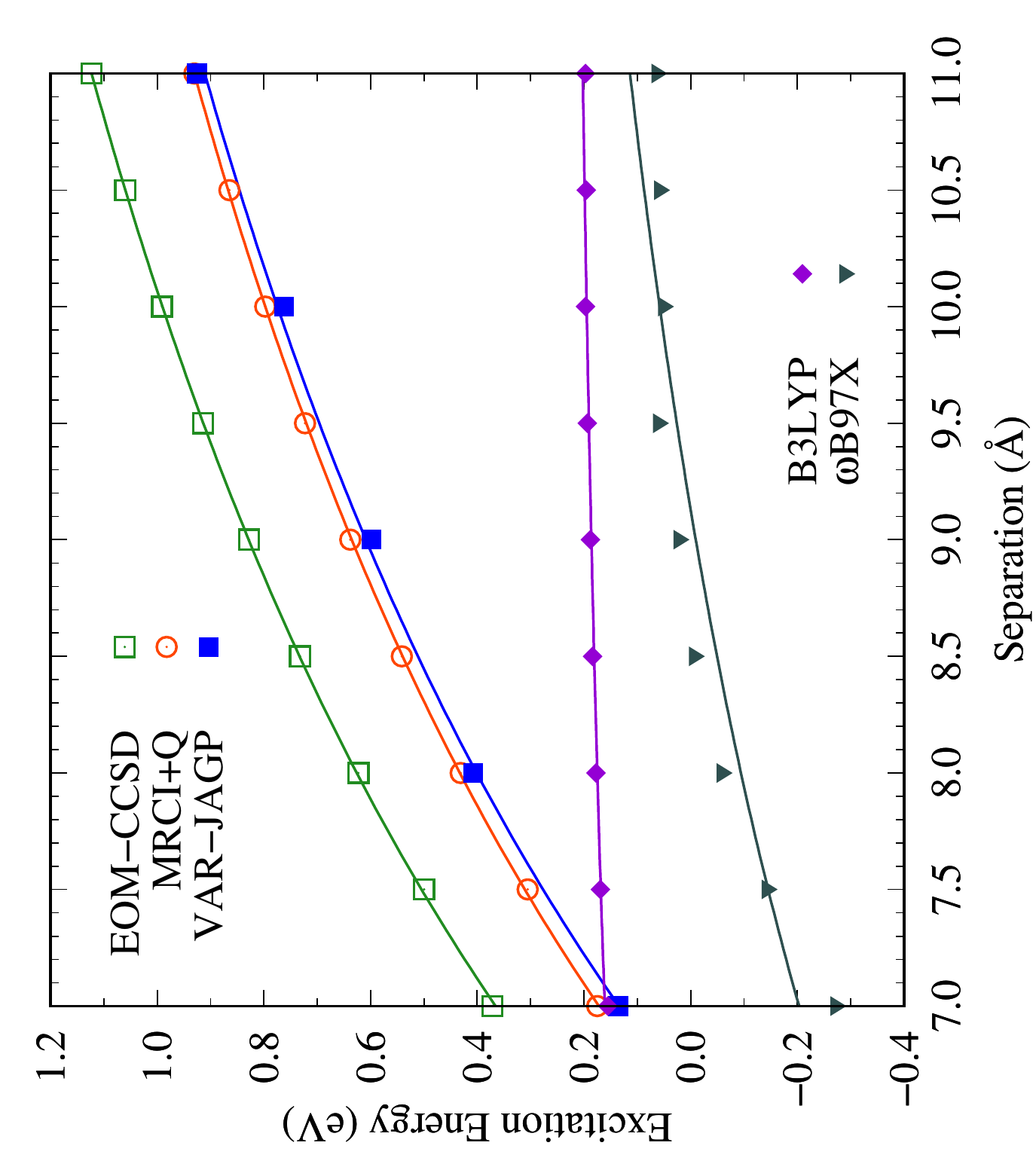}
\caption{Energy of the singlet-spin NaCl charge transfer excitation in the 6-31G basis.
         The lines are fits to $a-b/r$.
         The EOM-CCSD, MRCI+Q, and VAR-JAGP fits all have the correct $b=1.0$ $\mathrm{E_h}/$Bohr
         long range behavior for charge separation,
         while the B3LYP and $\omega$B97X fits have unphysical $b$ values of $0.06$ and $0.42$
         $\mathrm{E_h}/$Bohr, respectively.
        }
\label{fig:nacl_ct}
\end{figure}

Charge transfer excitations present serious challenges for TDDFT \cite{Head-Gordon:2003:tddft_ct} and CIS \cite{Subotnik:2011:cis_ct_bias},
meaning that quantitative accuracy for this chemically and technologically important class of excitations requires the use of higher-level
methods like EOM-CCSD.
As ground state QMC can, thanks to its lower scaling and better parallelism, be applied to system sizes beyond the reach of traditional
coupled cluster \cite{Alfe:2012:water_dft_dmc,Alfe:2013:water_dmc}, it is important to investigate the prospects for QMC-based excited state methods to treat charge transfer.
As a prototypical test, we have therefore applied VAR-JAGP to the low-lying charge transfer excitation in NaCl.
As seen in Figure \ref{fig:nacl_ct}, VAR-JAGP excels in this role, agreeing with the benchmark Davidson-corrected 
multi-reference configuration interaction (MRCI+Q) \cite{Werner:1988:mrci,Knowles:1988:mrci} result even more closely than the highly accurate EOM-CCSD.
For comparison, we also show TDDFT results for two functionals: the hybrid B3LYP \cite{BECKE:1993:b3lyp} and range-separated $\omega$B97X
\cite{Head-Gordon:2008:wb97x}.
While the latter provides for at least some distance dependence in the excitation energy, it is in this case unable to reproduce the
correct $1/r$ behavior that it is designed to capture due to self-interaction error in the restricted Khon-Sham representation of the Na$^{+}$Cl$^{-}$ state
that causes the HOMO to delocalize slightly onto the Na atom.
In Hartree-Fock, and thus also EOM-CCSD, the HOMO for this state is instead wholly located on Cl, as is also the case for the corresponding
natural orbital in JAGP.

\begin{figure}[t]
\includegraphics[width=8.0cm,angle=270]{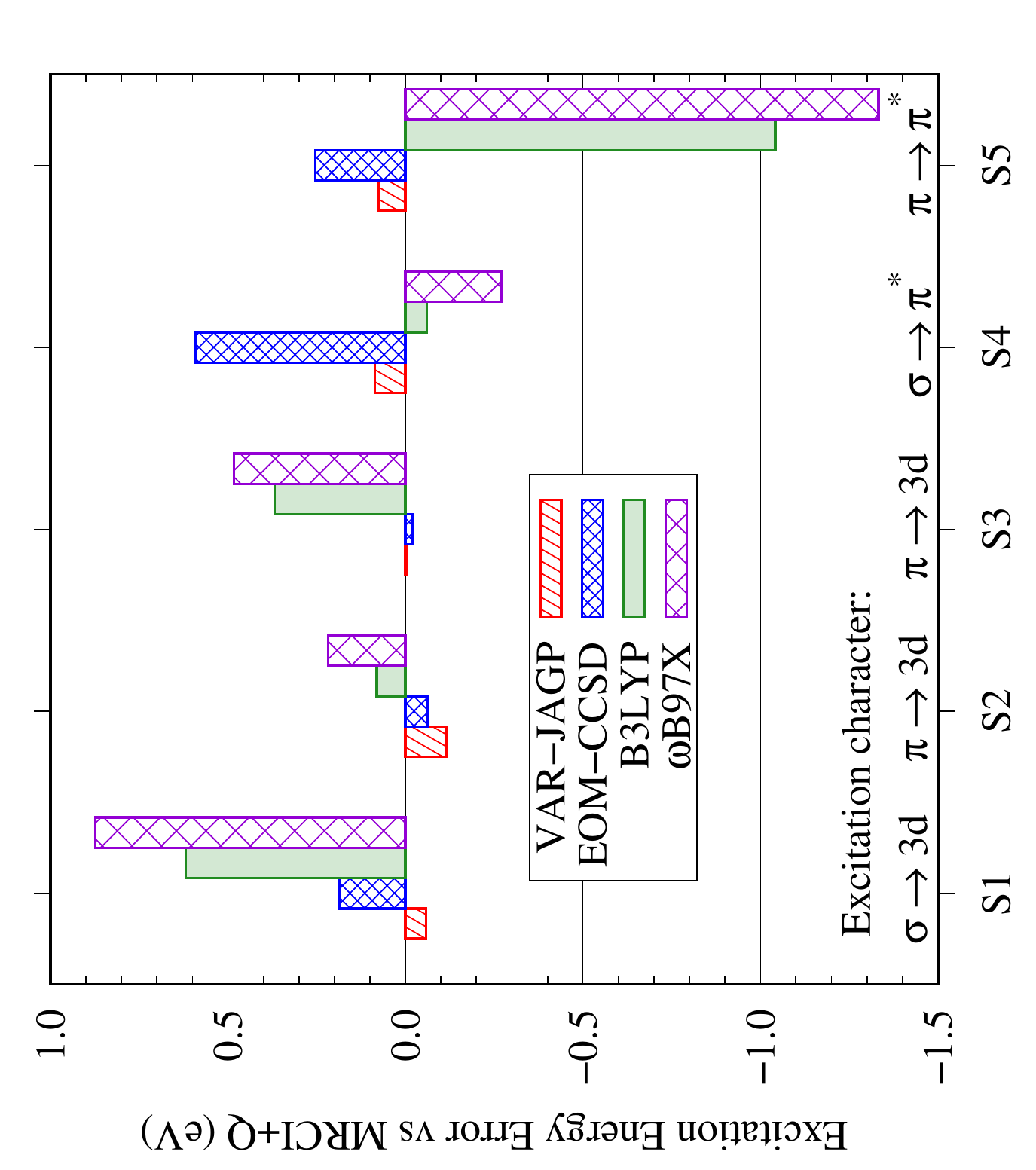}
\caption{Excitation energy errors for the first five singlet excitations in
         [VO]$^{3+}$ in a 6-31G basis.
         Note that in MRCI+Q, each of these excitations has at least 20\% of its weight
         on doubly or higher excited configurations.
        }
\label{fig:vo_error}
\end{figure}

As a final demonstration we have investigated the [VO]$^{3+}$ dimer, whose ground state is dominated by a closed shell configuration
featuring one $\sigma$ and two $\pi$ bonds, the latter of which have their density concentrated towards the oxygen.
Excitations from these bonds into either the nonbonding 3d or the $\pi^{*}$ orbitals (which are concentrated toward the metal)
will therefore involve varying degrees of short range charge transfer.
Indeed, this system was chosen as an example in which both charge transfer and significant double excitation character
are present simultaneously.
As seen in Figure \ref{fig:vo_error}, VAR-JAGP is the only method tested that reliably reproduced the excitation energies of
the MRCI+Q benchmark, having a worst-case error of just 0.11 eV.
While charge transfer alone should be handled well by EOM-CCSD, the presence of substantial doubly excited character in each of
these mostly single-excitation transitions leads to more sizable errors.
Hybrid-functional TDDFT is even less reliable, and the introduction of range separation does not improve matters, perhaps because
the charge transfers are occurring over such a short range that a formal separation between short and long range exchange is not appropriate.
As a final note, consider that the size of the LR space for JAGP is greatly smaller than that of CCSD (it is quadratic rather than quartic
in the system size), and yet VAR-JAGP proves significantly more accurate in this difficult case.
Our best explanation for this difference is the ability of VAR to relax the ground state wave function in the presence of the response,
an important flexibility that is lacking in traditional LR methods like CIS, EOM-CCSD, and TDDFT.

We have presented a new Monte Carlo approach for the coupled, variational optimization of an ansatz and its linear response for electronically
excited states.
Through explicit numerical demonstrations, we have shown that this coupled optimization captures the wave function relaxation effects
that are often missing in linear response approaches, such as the relaxation of the overall charge density in the presence of an excitation.
In initial testing, this variation-after-response approach displays an accuracy comparable to that of EOM-CCSD for single excitations, exceeds its
accuracy for double excitations, and performs very well for both short and long range charge transfer excitations.
Given that the approach is compatible with a wide variety of ansatzes and both open and periodic boundary conditions, we look forward
to further investigations of its usefulness in both chemistry and physics.

We acknowledge funding from the Office of Science, Office of Basic Energy Sciences, the US Department of Energy, Contract No. DE-AC02-05CH11231.
Calculations were performed on the Berkeley Research Computing Savio cluster.

\vspace{2mm}
\appendix
\noindent
Appendix:
\vspace{2mm}

EOM-CCSD and MRCI+Q results were computed with \uppercase{MOLPRO} \cite{MOLPRO_brief},
CIS and TDDFT results with QChem \cite{QChem:2006,QChem:2013} using the Tamm-Dancoff approximation,
and VAR-JAGP results with our own prototype Hilbert space
quantum Monte Carlo code with one- and two-electron integrals imported from PySCF \cite{pyscf}.
In VAR-JAGP we set our sample length as $2.88\times 10^6$ and worked exclusively in the symmetrically orthogonalized
``$S^{-1/2}$'' one particle basis.
Our geometries were given by rOH = 1.0\AA, $\angle$HOH = 109.57$^\circ$, rCC = 1.242514640\AA,
and rVO = 1.55\AA.
All calculations were performed in a 6-31G basis \cite{POPLE:1972:6-31g_basis} with the following
frozen cores.

\vspace{2mm}
\begin{tabular}{c c c}
\hline
  \hspace{0mm} Atoms & Methods & Frozen Cores \\
\hline
  \hspace{0mm} all & \hspace{5mm} TDDFT \hspace{5mm} & None \\
  \hspace{0mm} O, C  & \hspace{5mm} non-DFT methods \hspace{5mm} & He \\
  \hspace{0mm} Na, Cl & \hspace{5mm} EOM-CCSD, MRCI+Q \hspace{5mm} & He \\
  \hspace{0mm} V  & \hspace{5mm} EOM-CCSD, MRCI+Q \hspace{5mm} & Ne \\
  \hspace{0mm} Na, Cl & \hspace{5mm} VAR-JAGP \hspace{5mm} & Ne \\
  \hspace{0mm} V  & \hspace{5mm} VAR-JAGP \hspace{5mm} & Ar \\
\hline
\end{tabular}
\vspace{2mm}

\noindent
For VAR-JAGP, we fixed at zero the variable shifts $\vec{\mu}$ for Jastrow factor variables
to simplify the nonlinear optimization.
Note that were these not fixed, the optimization would have been more difficult due to greater linear dependency
in the variable space, but that the final results for each individual state would have been superior in a variational 
sense as guaranteed by the variational nature of our approach.

%

\bibliographystyle{aip}
\bibliography{var_qmc}

\end{document}